\documentclass[aps,gbroupedaddress,amsmath,amssymb]{revtex4-2}

\usepackage{anyfontsize} 

\usepackage {enumerate}
\usepackage {float}
\usepackage{epstopdf}
\usepackage{xcolor}

\usepackage{ulem}
\usepackage[breaklinks=true]{hyperref}
\usepackage{setspace}

\usepackage{amssymb}
\usepackage{stmaryrd}
\usepackage{amsmath}
\usepackage{amsfonts}
\usepackage{mathrsfs}
\usepackage[utf8]{inputenc}
\usepackage{amsmath,amssymb,amsfonts}
\usepackage{graphicx}
\usepackage{subfigure}
\usepackage{color} 
\usepackage{fancyhdr}
\usepackage{hyperref} 
\begin{document}
	\title[]{Gravitational waves for eccentric extreme mass ratio inspirals of self-dual spacetime}
	
	\author{Yunlong Liu}
	\affiliation{Department of Physics, South China University of Technology, Guangzhou 510641, China}
	\author{ Xiangdong Zhang\footnote{Corresponding author. scxdzhang@scut.edu.cn}}
	\affiliation{Department of Physics, South China University of Technology, Guangzhou 510641, China}


	\begin{abstract}
		In this paper, we calculate the frequencies of geodesic orbits in self-dual spacetime on the equatorial plane and obtain the leading-order effects of loop quantum parameters $P$ on the energy flux and angular momentum flux in eccentric extreme mass ratio inspirals. The gravitational waveform under different eccentricity is carried out by improved ``analytic-kludge'' method. Through the calculation of waveform mismatches for the LISA detector, the constraints on loop quantum parameters will be improved by 1 to 2 orders of magnitude, compared to the weak field experiments in the solar system, and can reach the level of $10^{-8}$.
	\end{abstract}

	\maketitle

	\section{Introduction}\label{Intro}
	General relativity (GR) has been developed for over a century, and so far the observed data have not exceeded the predictions of general relativity. However, there are irreconcilable problems between GR and quantum mechanics (QM). Hence, establishing a unified theory of GR and QM is the primary task in theoretical physics. Today, various quantum gravity theories have been proposed. Among them, loop quantum gravity (LQG) is notable due to its background independence and non-perturbative nature \cite{Quantum_Rovelli_2004, Modern_Thiemann_2007, Background_Ashtekar_2004, Fundamental_Han_2007}.

	The application of LQG to cosmology, known as loop quantum cosmology (LQC), now becomes a fruitful field. The most notable feature of LQC is that it successfully uses the ``Big Bounce'' to replace the classical big bang singularity \cite{Bojowald 2001, Ashtekar 2003, Modesto 2004, Quantum_Ashtekar_2005,Bojowald 2008, Ashtekar 2011}. 
 
	Due to the success of LQC, and note that the interior of Schwarzschild black hole is isometric to the Kantowski-Sachs model \cite{Quantum_Ashtekar_2005, Boehmer 2007}, the attempt to apply quantization techniques to black holes to address the issue of black hole singularity is a very intriguing idea. By utilizing the techniques developed in LQC and through different quantization schemes, various LQG black hole models can be constructed \cite{Loop_Zhang_2023}. Among these different models, one famous and well studied method is to set the loop quantum regularization parameters as constants\cite{Quantum_Ashtekar_2005, Semiclassical_modesto_2010}. By using the technique of holonomy correction, a so-called self-dual solution of loop quantum black holes without curvature singularities has been constructed \cite{Loop_Modesto_2006, Semiclassical_Modesto_2010}. 
 
	One natural question for a quantum black hole model is that does LQG parameters can generate any observable features, or whether these quantum parameters can be constrained through current or future experiments and observations? A lot of effects have been made towards this direction. For example, in the solar system \cite{Confrontation_will_2014}, or in strong field regimes  observations of binary pulsars \cite{Binary_taylor_1994, Doublebinarypulsar_yunes_2009, Testing_seymour_2018}, imaging of black hole shadows \cite{First_EHT_2019, Gravitational_ehtcollaboration_2020}, and direct detection of gravitational wave(GW) \cite{Tests_ligo_2016}. Based on this, lots of properties and observational constraints in self-dual spacetimes have been studied \cite{Gravitational_Sahu_2015,Observational_Zhu_2020,Constraints_Yan_2022, Constraints_Liu_2023}. As a result, the best constraints on LQG parameters are provided within the solar system experiments \cite{Observational_Zhu_2020}, with a limit of quantum parameter $|\delta| < 0.0199$ (or $P<5.5 \times 10^{-6}$).
	
	To obtain with more tighter constraints on parameters, more precise experimental tests are necessary. The detection of GWs presents an opportunity for this. The next generation of space-based detectors, such as LISA \cite{Laser_Amaro-Seoane_2017}, Tianqin \cite{Science_Fan_2020}, and Taiji \cite{Taiji_Hu_2017,Alternative_Wang_2021}, will expand the types and spectra of GW signals to unprecedented precision, providing possibilities to testing LQG and constrain the quantum parameters.
	
	Among various GW signals, an important one comes from the extreme mass ratio inspiral (EMRI) system composed of a massive black hole ($M \approx 10^{5 \sim 7} M_\odot$) and a small compact object ($m \approx 10^{0 \sim 2} M_\odot$) \cite{Science_Babak_2017}. Analyzing EMRIs requires the establishment of corresponding waveform models \cite{LISA_Barack_2004,Kludge_Babak_2007,Improved_Chua_2015, Augmented_Chua_2017, Fast_Katz_2021,Fast_Speri_2023}. For example, the high-precision kludge method \cite{Kludge_Babak_2007} has been proposed. However, these methods come with a high computational cost. To save computational resources, Barack and Cutler \cite{LISA_Barack_2004} proposed the ``analytic kludge'' (AK) model. 
	This model can quickly obtain a GW template of EMRI, and can reflect the main physical properties of the real EMRI waveform\cite{Kludge_Babak_2007}. However, when this method approaches the innermost stable orbit (ISO), the phase error will increase rapidly. In order to solve this problem, Chua et al. proposed an improved AK waveform method \cite{Improved_Chua_2015,Augmented_Chua_2017} with ``map'' way, and obtained more accurate waveforms.

	In this paper, we propose a new improvement on the AK method differs from the ``map'' way of Chua \cite{Improved_Chua_2015,Augmented_Chua_2017}. We used the inverse series way and obtain the high precision perihelion precession angle, which encodes the major cumulative orbital error in spherically symmetric spacetime when energy and angular momentum fluxes of GWs are not considered. As a result, we calculate the impact of LQG parameter on GWs and explore the detection capabilities of space-based GW detectors for the parameter of LQG. We find that the constraint ability on loop quantum parameters will be greatly improved  compared with the  experiments in the solar system.
	
	The structure of this paper is as follows:
	In Sec. \ref{SandW}, we introduce the geodesic equations on the self-dual  spacetime and  calculate two mean orbital frequencies on the equatorial plane, and give the leading-order correction of fluxes due to the LQG effects.
	In Sec. \ref{ODE}, we provide two improved adiabatic evolution equations for orbitals and also present the quadrupole moment waveform in the self-dual spacetime. Then, in Sec. \ref{GW}, we obtain the numerical solutions of  ODEs from the previous section and corresponding gravitational waveforms. With all this, we analyze the detection capability of LISA for the parameter of LQG with mismatch in Sec. \ref{PE}.
	Finally, the conclusions and perspectives are provided in Sec. \ref{conclusion}.

	\section{The background metric and motion equation}\label{SandW}

	\subsection{background metric}\label{spacetime}
	
	The metric of static spherically symmetric self-dual black hole  is given by\cite{Semiclassical_Modesto_2010,Loop_Zhang_2023}:
	\begin{eqnarray}
		d s^{2}= - A(r) dt^2 +  B(r) dr^2 +  D(r) d\Omega^2,
	\end{eqnarray}
	where
	\begin{eqnarray}
		A(r) &=& \frac{\left(r-r_{+}\right)\left(r-r_{-}\right)\left(r-r_{*}\right)}{r^{4}+a_{0}^{2}},\\
		B(r) &=& \frac{\left(r+r_{+}\right)^{2}\left(r^{4}+a_{0}^{2}\right)}{\left(r-r_{+}\right)\left(r-r_{-}\right) r^{4}}, \\
		D(r) &=&r^{2}+\frac{a_{0}^{2}}{r^{2}}.
	\end{eqnarray}
	Here, $r_{+}=2M/(1+P)^2$, $r_{-}=2MP^2/(1+P)^2$, and $r_{*}=\sqrt{r_{+}r_{-}}$. 
	It should be noted that $a_{0}$  depends on the minimal area element $\Delta$ and $P$ corresponds to the regularization parameter $\delta$ in LQG as
	\begin{eqnarray}
		a_{0}=\frac{\Delta}{8 \pi}; \quad P=\frac{\sqrt{1+\gamma^{2} \delta^{2}}-1}{\sqrt{1+\gamma^{2} \delta^{2}}+1};
	\end{eqnarray}
	When $a_0=0$ and $P=0$, the spacetime can return to the Schwarzschild case. Since the parameter $a_0$ usually appears at higher order terms in the Taylor series expansion while the gravitational wave we concern in this paper only involves quadrupole which is irrelevant with higher order terms and hence we can safely set $a_0=0$ in the following. Moreover, we adopt geometric units $G=c=1$.
	
	\subsection{Equations of motion in the spherically symmetric metric}\label{wavefun}
	Now we turn to the equations of motion. By using the Killing vectors, we can define two conserved quantities, energy $E$ and angular momentum $L$ as
	\begin{eqnarray}
		E &:=&  m  A(r)\frac{dt}{d\tau}, \\
		L &:=&  m  \sin^2(\theta) D(r)\frac{d\phi}{d\tau}.
	\end{eqnarray}
	Through the Lagrangian and separating variables, we can obtain the equations of motion of this spacetime as follows:
	\begin{eqnarray}
		&&\frac{dt}{d\tau}=\frac{E}{m A(r)},  \label{dtdtau}\\
		&&(\frac{dr}{d\tau})^2=\frac{B(r)}{D(r)}\left(-D(r)+\frac{D(r)}{A(r)} \frac{E^{2}}{m^{2}}-\frac{C}{m^{2}}\right), \label{drdtau}\\
		&&(\frac{d\theta}{d\tau})^2=\frac{C}{m^{2}}-\frac{L^{2}}{m^{2} \sin^{2}(\theta)}, \label{dthetadtau} \\
		&&\frac{d\phi}{d\tau}=\frac{L}{m \sin^2(\theta) D(r)}. \label{dphidtau}
	\end{eqnarray}
	
	Moreover, to facilitate calculations, we 
	perform a variable transformation from the original coordinate system $(t, r, \theta, \phi)$ to the new one $(t, \psi, \theta, \phi)$, where
	\begin{eqnarray} \label{rtopsi}
		r(\psi)= \frac{M p}{1 +  e \cos(\psi)},
	\end{eqnarray}
	where $p$ is the dimensionless semilatus rectum and $e$ is the eccentricity.

	Note that the distances of the orbit's pericenter and apocenter are respectively given by:
	\begin{eqnarray}
		&&r_{\text{apo}}= \frac{M p}{1-e}, \\ 
		&&r_{\text{peri}}=\frac{M p}{1+e}.
	\end{eqnarray} For $0<e<1$, the radial velocity of the small body equal zero ($dr/d\tau=0$) at the pericenter or apocenter, where $\psi=\mathbb{N} \pi, \mathbb{N}=1,2,3,...$. 
	Hence, $dr/d\tau$ and $\sin(\psi)$  change sign on  $\psi=\mathbb{N} \pi$ at the same time. 
	Take the derivative of Eq. \eqref{rtopsi}, we get
	\begin{eqnarray} \label{dpsidtau}
	 \frac{d\psi}{d\tau} = \frac{(1 +  e \cos(\psi))^2}{e M p}\frac{1}{ \sin(\psi)}\frac{dr}{d\tau}. 
	\end{eqnarray} Therefore, $d\psi/d\tau$ always retains the same sign. This feature enables us to set $d\psi/d\tau \geq 0$ which indicates $\psi$ monotonically increases with proper time $\tau$.
	For equatorial plane orbits, we have  $\theta=\pi/2$ and $d\theta/d\tau=0$ . From Eq. \eqref{dthetadtau}, we have $C=L^2$. 
	The equations of motion can be reduced to:
	\begin{eqnarray}
		&&\frac{dt}{d\tau}=\frac{E}{m A(r(\psi))}, \label{EquatorialEqT}\\
		&&\frac{d\psi}{d\tau}=\frac{(1+e \cos(\psi))^{2}}{e M p \sqrt{\sin(\psi)^{2}}} 
		\sqrt{\frac{1}{B(r(\psi))}\left(- 1+\frac{1}{A(r(\psi))} \frac{E^{2}}{m^{2}}-\frac{1}{D(r(\psi))}\frac{L^{2}}{m^{2}}\right)}, \label{EquatorialEqR}\\
		&&\frac{d\theta}{d\tau}=0 , \\
		&&\frac{d\phi}{d\tau}=\frac{L}{m D(r(\psi))}. \label{EquatorialEqPhi}
	\end{eqnarray}
	Next, let's further analyze the relationship between $(p,e)$ and $(E,L)$.
	Since we have $dr/d\tau=0$ at $r_{\text{apo}}$ and $r_{\text{peri}}$, we can obtain the relationship between $(p,e)$ and $(E,L)$ from Eq. \eqref{dthetadtau} as follows: 
	\begin{eqnarray}
	E = \frac{m \sqrt{A(r_{\text{apo}})} \sqrt{A(r_{\text{peri}})} \sqrt{D(r_{\text{apo}})-D(r_{\text{peri}})}}{\sqrt{A(r_{\text{peri}}) D(r_{\text{apo}})-A(r_{\text{apo}}) D(r_{\text{peri}})}}, \label{Etope}\\ 
	L = \frac{m \sqrt{A(r_{\text{apo}})-A(r_{\text{peri}})} \sqrt{D(r_{\text{apo}})} \sqrt{D(r_{\text{peri}})}}{\sqrt{A(r_{\text{peri}}) D(r_{\text{apo}})-A(r_{\text{apo}}) D(r_{\text{peri}})}}. \label{Ltope}
	\end{eqnarray}
	It should be note that when $e=0$, Eqs. \eqref{dpsidtau} and \eqref{EquatorialEqR} are not valid, but we can do a similar analysis. Since in this case, we have $dr/d\tau=0$, $dr^2/d\tau^2=0$ and $r_{\text{apo}}=r_{\text{peri}}= M p$, hence we can obtain:
	\begin{eqnarray}
		E=\frac{m A\left[r_{\text {apo }}\right] \sqrt{D^{\prime}\left[r_{\text {apo }}\right]}}{\sqrt{-D\left[r_{\text {apo }}\right] A^{\prime}\left[r_{\text {apo }}\right]+A\left[r_{\text {apo }}\right] D^{\prime}\left[r_{a p o}\right]}}, \\ L=\frac{m D\left[r_{\text {apo }}\right] \sqrt{A^{\prime}\left[r_{\text {apo }}\right]}}{\sqrt{-D\left[r_{\text {apo }}\right] A^{\prime}\left[r_{\text {apo }}\right]+A\left[r_{\text {apo }}\right] D^{\prime}\left[r_{\text {apo }}\right]}}.
	\end{eqnarray}
	Expanding the energy $E$ and angular momentum $L$ in terms of the dimensionless semilatus rectum $p$, we have:
	\begin{eqnarray}
		\frac{E}{m} &=&1 + \frac{(-1 + e^2) (1-P)^2}{2 (1 + P)^2 p} + \frac{3 (-1 + e^2)^2 (1-P)^4}{8 (1 + P)^4 p^2} + \mathcal{O}(p^{-3}),  \label{ETop}\\ 
		\frac{L}{Mm} &=&\frac{1 - P}{1 + P}p^{1/2} -  \frac{3 + e^2 (1-P)^4 - 4 P + 10 P^2 - 4 P^3 + 3 P^4}{2 (1-P) (1 + P)^3} p^{-1/2} + \mathcal{O}(p^{- 3/2}). \label{LTop}
	\end{eqnarray}
	It is easy to see that when $P=0$, the above result reduced to the Schwarzschild case, while $e=0$, the case of circular orbit will be recovered.
	
	\subsection{Orbital frequencies}\label{OrbitF}
	
	We assume that the trajectory is on the equatorial plane $\theta=\pi/2$, it means that the polar angle frequency $\Omega_\theta$ can be neglected. For bounded orbits, we are left with  two frequencies $\Omega_r$ and $\Omega_\phi$  that correspond to the parameters $r(\psi)$ and $\phi$.  
	The coordinate time spent by the test body in moving from one pericenter to the next one is given by \cite{Analytical_Fujita_2009}:
	\begin{eqnarray}
		\Lambda_t=\int_0^{t_0} dt = \int_0^{2\pi} \frac{dt}{d\psi} d\psi.
	\end{eqnarray}
	The corresponding change in the azimuthal angle $\phi$ reads:
	\begin{eqnarray}
		\Lambda_\phi=\int_0^{\phi_0} d\phi = \int_0^{2\pi} \frac{d\phi}{d\psi} d\psi.
	\end{eqnarray}
	For this, the average angular velocity of $r(\psi)$ and $\phi$ over each period can be expressed as
	\begin{eqnarray}
		\Omega_r= \frac{2 \pi}{\Lambda_t};  \quad	\Omega_\phi =   \frac{ \Lambda_\phi }{ \Lambda_t}.
	\end{eqnarray}
	By combining the geodesic equations \eqref{EquatorialEqT} and \eqref{EquatorialEqPhi}, we can obtain the following expansion:
	\begin{eqnarray}
		\Omega_\phi &&=\frac{(1-P) X^{3/2}}{(1 + P) M p^{3/2}} -\frac{X^{3/2}}{(1-P) (1 + P)^3 M p^{5/2}}\notag\\
		&&\times \Bigl( -3 + 4 P - 6 P^2 + 4 P^3 - 3 P^4 + (3 - 8 P + 10 P^2 - 8 P^3 + 3 P^4) X \Bigr) \notag\\
		&&-\frac{X^{3/2}}{4 (1-P)^3 (1 + P)^5 M p^{7/2}}\Bigl(-57 - 4 (1-P)^4 X^2 (3 + \mathcal{P}_{2}{})^2  \notag\\ 
		&&+ 2 (1-P)^4 X^{3/2} (15 + \mathcal{P}_{4}{}) + (1-P)^2 X (63 + \mathcal{P}_{6}) + \mathcal{P}_{8}\Bigr) + \mathcal{O}(p^{- 9/2}) ,   \label{Omegar}\\
		\Omega_r &&= \frac{(1-P) X^{3/2}}{(1 + P) M p^{3/2}} - \frac{(1-P) X^{5/2} (3 + \mathcal{P}_{2}{})}{(1 + P)^3 M p^{5/2}}   \notag\\
		&&-\frac{(1-P) X^{5/2} }{2 (1 + P)^5 M p^{7/2}}
		\left(12 (1 + 3 P^2 + P^4) - 2 X (3 + \mathcal{P}_{2}{})^2 + X^{1/2} (15 + \mathcal{P}_{4}{})\right) + \mathcal{O}(p^{- 9/2}), \label{Omegaphi}
	\end{eqnarray}
	where
	\begin{eqnarray}
		\mathcal{P}_2= &&-2 P + 3 P^2 ,\\
		\mathcal{P}_4= &&-12 P + 26 P^2 - 12 P^3 + 15 P^4,\\
		\mathcal{P}_6= &&-122 P + 233 P^2 - 284 P^3 + 233 P^4 - 122 P^5 + 63 P^6,\\
		\mathcal{P}_8= &&200 P - 364 P^2 + 568 P^3 - 662 P^4 + 568 P^5 - 364 P^6 + 200 P^7 - 57 P^8.
	\end{eqnarray}
	Thus, we can obtain the frequency of perihelion precession $\dot{\hat{\gamma}}$ as:
	\begin{eqnarray} 
		\dot{\hat{\gamma}}&= &\frac{(-3 + 4 P - 6 P^2 + 4 P^3 - 3 P^4) X^{3/2} - (1 - P) (2 P - 5 P^2 + 3 P^3 + (1-P)\mathcal{P}_{2})X^{5/2}}{-(1 - P) (1 + P)^3 M p^{5/2}} \notag\\
		&&+ \frac{X^{5/2} \bigl(-24 (1 - P)^4 (1 + 3 P^2 + P^4) + (1 -  P)^2 (63 + \mathcal{P}_{6})\bigr) + X^{3/2} (-57 + \mathcal{P}_{8})}{-4 (1 - P)^3 (1 + P)^5 M p^{7/2}} + \mathcal{O}(p^{- 9/2}), \label{gammaF}
	\end{eqnarray}
	where $X=1-e^2$.
	Here, it can be easily verified that when $P=0$, the result returns to the Schwarzschild case. In our code, in order to obtain a higher precision geodesic orbit, we use a higher order series expansion formula of  $\dot{\hat{\gamma}}$. However, due to the increased complexity, we will not show it here.
	
	\subsection{Fluxes}\label{result}
	In the AK waveform on the equatorial plane, in addition to the two motion equations obtained from geodesics, there are also adiabatic evolution for frequency $\nu$ and eccentricity $e$ caused by energy and angular momentum flux of GW. 
	
	We use the quadrupole approximation to calculate the energy flux and angular momentum flux\cite{Multipole_Thorne_1980, Gravitational_Maggiore_2007} as
	\begin{eqnarray}
		\frac{d E}{d t}&=&\frac{1}{5}\left\langle\dddot{\mathcal{Q}}_{i j} \dddot{\mathcal{Q}}_{i j}\right\rangle,\\
		\frac{d L^{i}}{d t}&=&\frac{2}{5} \epsilon^{i k l}\left\langle\ddot{\mathcal{Q}}_{k a} \dddot{\mathcal{Q}}_{l a}\right\rangle,
	\end{eqnarray}
	where the quadrupole moment $\mathcal{Q}$ can be described in terms of the mass moment $\mathcal{M}$ as follows:
	\begin{eqnarray}
	\mathcal{Q}^{i j}&=& \mathcal{M}^{i j}-\frac{1}{3} \delta^{i j} \mathcal{M}_{k k}, \label{QM}\\
	\mathcal{M}^{i j}&=&\int d^{3} x T^{00}(t, \mathbf{x}) x^{i} x^{j},
	\end{eqnarray}
	where $T^{ab}$ is the stress-energy tensor and $\mathbf{x}$ is the position vector. In the point particle approximation and in the center-of-mass coordinates, the second mass moment can be expressed as:
	\begin{eqnarray}
		\mathcal{M}^{ij}=\mu x_0^i x_0^j,
	\end{eqnarray}
	where $\mu=mM/(m+M)$ is the reduced mass, and $\mathbf{x}_0$ is the relative position between the small body and the central black hole. 
	In the weak field approximation, $x_0^i = (r \cos \phi, r \sin \phi, 0)$, the leading-order energy flux and angular momentum flux are given as follows:
	\begin{eqnarray}
		\frac{1}{m} \frac{dE}{dt}&=&- \frac{(1 -  e^2)^{3/2} (96 + 292 e^2 + 37 e^4) (-1 + P)^6 \mu^2}{15 (1 + P)^6 M^2 m p^5} + \mathcal{O}(p^{-6}), \label{dEdt}\\
		\frac{1}{m M}\frac{dL}{dt}&=&- \frac{4 (1 -  e^2)^{3/2} (8 + 7 e^2) (1 -  P)^5 \mu^2}{5 (1 + P)^5 M^2 m p^{7/2}}+ \mathcal{O}(p^{- 9/2}). \label{dLdt}
	\end{eqnarray}

	\section{Orbital evolution equations}\label{ODE}
	
	Now we have the expressions for energy flux and angular momentum flux in terms of a series expansion in Eq.\eqref{dEdt}-\eqref{dLdt}, the next step is to convert this series expansion in terms of the Keplerian frequency $\nu$. 
	Additionally, in equations \eqref{Omegar} and \eqref{Omegaphi}, there are two frequencies involved, $\Omega_r$ and $\Omega_\phi$. In the conventional approach of the $AK$ method, it is usually assumed that $\Omega_r \approx 2\pi\nu$ \cite{Improved_Chua_2015, Augmented_Chua_2017}. However, this approximation introduces significant errors, especially when approaching the innermost stable orbit(ISO) \cite{Improved_Chua_2015, Augmented_Chua_2017}. A detailed comparison of $\Omega_r$, $\Omega_\phi$ and $2\pi\nu$ is provided in Appendix \ref{FProblem}. Since this paper only considers the spherically symmetric case, we do not use the ``map'' method proposed in \cite{Improved_Chua_2015} to address the large discrepancies between the orbital and precession frequency $\{f_{orb}, f_{peri}\}$ and the approximations of $\{ \Omega_r/2\pi, (\Omega_\phi-\Omega_r)/2\pi \}$. Instead, we will directly employ higher order series expansions of $\dot{\hat{\gamma}}$ to resolve this issue.
	
	\subsection{Geodesic Equations}
	In this part, we will use both frequencies, $\Omega_r= 2 \pi \nu_r$ and $\Omega_\phi= 2 \pi \nu_\phi$, as base frequencies to obtain the approximate equations of geodesics.
	
	\subsubsection{The large eccentricity method (eL method)}
	We assume $\Omega_r= 2 \pi \nu_r$, and by calculating the inverse series expansion of Eq. \eqref{Omegar}, we obtain the relationship between $p$ and the radial frequency $\nu_r$:
	\begin{eqnarray}
		p &=& \left(\frac{1-P}{1+P}\right)^{2 / 3} \frac{X}{(2 \pi M \nu_r)^{2 / 3}}-\frac{2(3+P(-2+3 P)) X}{3(1+P)^{2}} \notag \\
		&&+(2 \pi M \nu_r)^{2 / 3}\left(\frac{1+P}{1-P}\right)^{2 / 3} \frac{1}{9(1+P)^{4}}  \Bigl(-36\left(1+3 P^{2}+P^{4}\right)\notag \\
		&&+3\left(-15+12 P-26 P^{2}+12 P^{3}-15 P^{4}\right) \sqrt{X}+\left(3-2 P+3 P^{2}\right)^{2} X\Bigr)+ \mathcal{O}((2 \pi M \nu_r)^{4/3}).  \label{pTonu_r}
	\end{eqnarray}
	If we substitute Eq. \eqref{pTonu_r} into Eq. \eqref{Omegar}, we can obtain $\Omega_r= 2 \pi \nu_r + \mathcal{O}(\nu_r^{7/3})$. By substituting the expression \eqref{pTonu_r} into Eq. \eqref{gammaF}, evolution equation determined by geodesic is given by
	\begin{eqnarray}
		\frac{d \Psi}{d t}&=&\Omega_r = 2 \pi \nu_r , \label{psifunel}\\
		\frac{d \Phi}{d t}&=& \Omega_r + \frac{d \hat{\gamma}}{d t}, \label{phiel}\\
		\frac{d \hat{\gamma}}{d t}&=&\frac{\left(1+P^{2}\right)\left(3-4 P+3 P^{2}\right)}{(1-P)^{8 / 3}(1+P)^{4 / 3}} \frac{(2 \pi M \nu_r)^{5 / 3}}{\left(1-e^{2}\right) M}\notag \\
		&&+\frac{1}{(1-P)^{16 / 3}(1+P)^{8 / 3}} \frac{(2 \pi M \nu_r)^{7 / 3}}{12 M\left(1-e^{2}\right)^{2}}\left(2\left(117+\mathscr{P}_{2\gamma 0}^{8}\right)- e^{2}\left(63+\mathscr{P}_{\gamma 2}^{8}\right)\right)+\mathcal{O}((2 \pi M \nu_r)^{3}),  \label{gammafunel}
	\end{eqnarray} 
	where $\Psi$ represents mean anomaly  for $\psi$ and $\Phi$ denotes mean ``longitude'' for  $\phi$  with
	\begin{eqnarray}
		\mathscr{P}_{2\gamma 0}^{8}&=&-432 P+860 P^{2}-1360 P^{3}+1582 P^{4}-1360 P^{5}+860 P^{6}-432 P^{7}+117 P^{8}, \\ 
		\mathscr{P}_{\gamma 2}^{8}&=&\left(7-6 P+7 P^{2}\right)\left(3-5 P+5 P^{2}-3 P^{3}\right)^{2}-63. 
	\end{eqnarray}	
	 
	\subsubsection{The small eccentricity method(eS method)}
	Analogously, we set $\Omega_\phi = 2 \pi \nu_\phi$. By computing the inverse series of Eq. \eqref{Omegaphi}, we can obtain the relation between  $p$ and the ``longitude'' frequency $\nu_\phi$:
	\begin{eqnarray}
		p &=&\left(\frac{1-P}{1+P}\right)^{2 / 3} \frac{X}{(2 \pi M \nu_\phi)^{2 / 3}} -\frac{2(3+P(-2+3 P)) X}{3(1+P)^{2}}+\frac{2\left(1+P^{2}\right)(3+P(-4+3 P))}{3\left(-1+P^{2}\right)^{2}}\notag\\
		&&+F[P,X] (2 \pi M \nu_\phi)^{2 / 3} +\mathcal{O}\left( (2 \pi M \nu_\phi)^{4/3}\right). \label{pTonu_phi}
	\end{eqnarray}
	Here, 
	\begin{eqnarray}
		F[P,X]&=&\frac{(1-P)^{10 / 3}}{9(1+P)^{10 / 3}} \Big({\left(81-360 P+572 P^{2}-984 P^{3}+1126 P^{4}-984 P^{5}+572 P^{6}-360 P^{7}+81 P^{8}\right)}X^{-1}+ \notag\\ 
		&&\left(-9+6 P+P^{2}+132 P^{3}+P^{4}+6 P^{5}-9 P^{6}\right)(1-P)^{2} \notag\\ 
		&&-6\left(15-12 P+26 P^{2}-12 P^{3}+15 P^{4}\right)(1-P)^{4} \sqrt{X}+2\left(3-2 P+3 P^{2}\right)^{2}(1-P)^{4} X \Big).
	\end{eqnarray}

	Substituting Eq.\eqref{pTonu_phi} into Eq. \eqref{Omegaphi}, we get $\Omega_\phi= 2 \pi \nu_\phi + \mathcal{O}(\nu_\phi^{7/3})$. By plugging the Eq. \eqref{pTonu_phi} into Eq. \eqref{gammaF}, equations of motion can be obtained as
	\begin{eqnarray}
		\frac{d \Psi}{d t}&=&\Omega_r =\Omega_\phi - \frac{d \hat{\gamma}}{d t}, \label{psifunes}\\
		\frac{d \Phi}{d t}&=& \Omega_\phi = 2 \pi \nu_\phi,  \label{phifunes}\\
		\frac{d \hat{\gamma}}{d t}&=&-\frac{\left(1+P^{2}\right)\left(3-4 P+3 P^{2}\right)}{(1-P)^{8 / 3}(1+P)^{4 / 3}} \frac{(2 \pi M \nu_\phi)^{5 / 3}}{\left(1-e^{2}\right) M}\notag \\
		&&-\frac{1}{(1-P)^{16 / 3}(1+P)^{8 / 3}} \frac{(2 \pi M \nu_\phi)^{7 / 3}}{12 M\left(1-e^{2}\right)^{2}}
		\left(2\left(27+\mathscr{P}_{1\gamma 0}^{8}\right)-e^{2}\left(63+\mathscr{P}_{\gamma 2}^{8}\right)\right)+\mathcal{O}((2 \pi M \nu_r)^{3}),  \label{gammafunes}
	\end{eqnarray}
	where 
	\begin{eqnarray}
		\mathscr{P}_{1\gamma 0}^{8}&=&-192 P+340 P^{2}-640 P^{3}+722 P^{4}-640 P^{5}+340 P^{6}-192 P^{7}+27 P^{8}. 
	\end{eqnarray}

	Along these lines, we can extand the approximate Eq. \eqref{gammafunel} or Eq. \eqref{gammafunes} to higher orders. In the following part of the paper, whenever we talk about using Eq. \eqref{gammafunel}  or Eq. \eqref{gammafunes}  up to $(n,m)$ orders, it means that the approximation of $d\hat{\gamma}/dt$ will be calculated to $n$-th orders ($\mathcal{O}((2 \pi M \nu_\kappa)^{(2n+3)/3}),\ \kappa=r, \phi$) in Schwarzschild part and $m$-th orders in LQG part. For example,the expressions in Eq. \eqref{gammafunel}  or Eq.\eqref{gammafunes} is up to $(2,2)$ orders. 
	
	\subsubsection{Orbits}
	As the extreme mass ratio parameter $m/M$ approaches zero, the energy and angular momentum flux also tend to zero, and the orbit can be described by a timelike geodesic and the information of orbit is completely encoded in the Eqs. \eqref{psifunel}-\eqref{gammafunel} or \eqref{psifunes}-\eqref{gammafunes}. 
	However, the azimuthal angle $\Psi$ and $\Phi$ are the mean of true angel $\psi$ and $\phi$. 
	The relationship between the true anomaly $\psi$ and the mean anomaly $\Psi$ can be derived using classical methods, and it is as follows\cite{Gravitational_Maggiore_2007}:
	\begin{eqnarray} \label{psiApprox}
		\psi&\approx&\psi_{Ap}=2 \arctan\left(\left(\frac{1+e}{1-e}\right)^{1/2}\tan\left(\frac{u}{2}\right)\right) + 2 \pi \mathrm{IntegerPart}(\frac{\Psi+\pi}{2 \pi}  ), \\
		u&=&\Psi+e \sin(u).
	\end{eqnarray}	
	The third order approximation for $u$ is:
	\begin{eqnarray}
		u=\Psi+e \sin(\Psi+e \sin((\Psi+e \sin (\Psi))).
	\end{eqnarray}	Moreover, $\phi$ can be approximated as:
		\begin{eqnarray}\label{phiApprox}
		\phi \approx \phi_{Ap} = \psi_{Ap} (1 + \frac{\hat{\gamma}}{\Psi} ).
	\end{eqnarray}	
	
	The approximate orbit is given by $r^i=(r \cos(\phi), r\sin(\phi),0)$, where $r$ is defined in Eq. \eqref{rtopsi}. In Eq. \eqref{phiApprox} , the use of the mean precession angle $\hat{\gamma}$ and mean anomaly $\Psi$ introduces periodic errors in the orbit.  
	In addition, from Eqs. \eqref{psifunel} - \eqref{gammafunel} or \eqref{psifunes} -\eqref{gammafunes}, we know that the main and cumulative error is mainly in  $\hat{\gamma}$. Hence, the analysis of the error of $\hat{\gamma}$ is helpful for us to estimate the orbit accuracy which well be shown in Appendix \ref{OrbitError}.
	
	Note that, to reduce this error accumulates over time, we usually employing higher orders of $d\hat{\gamma}/dt$ in Eq. \eqref{gammafunes} or 
	\eqref{gammafunel} when calculating numerical solutions. 
	By numerically solving the system of Eqs. \eqref{EquatorialEqT}-\eqref{EquatorialEqPhi}, Eqs. \eqref{psifunel}-\eqref{gammafunel} up to  $(10,5)$ orders, or Eqs. \eqref{psifunes} -\eqref{gammafunes} up to  $(11,6)$ orders, we obtain the following orbital results as shown in the Fig. \ref{SD_Geodesic}:
	
	\begin{figure}[!htb]
		\centering
		\subfigure[{$e=0.1$}]{
		\includegraphics[width=0.45\textwidth]{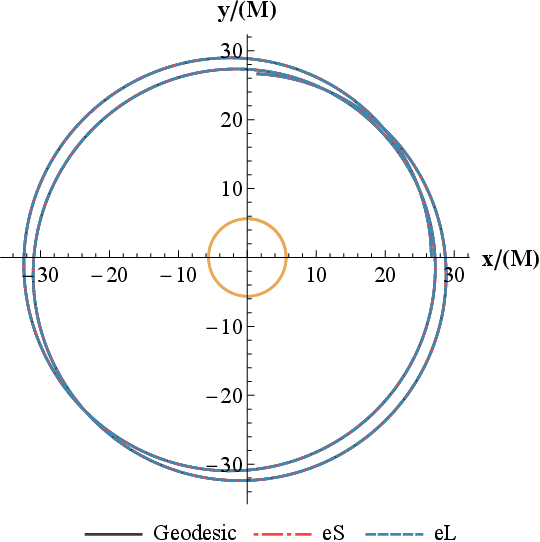}
		}
		\subfigure[{$e=0.8$}]{
		\includegraphics[width=0.45\textwidth]{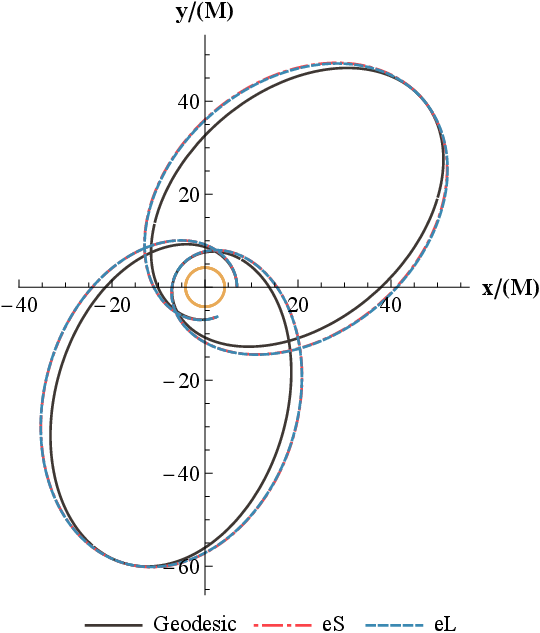}
		}
		\caption{ Equatorial orbits: the solid black line represents high precision numerical result for Eqs. \eqref{EquatorialEqT}-\eqref{EquatorialEqPhi}, while the dotted lines with different colors represent the approximate orbits obtained using eL and eS method. The mass of supermassive black hole is set to be $M=10^6 M_\odot$, and the loop quantum parameter is chosen as $P=1/1000$. The frequency of  $\phi$  is $\Omega_\phi/(2 \pi)=0.2 \mathrm{mHz}$, corresponding to an approximate orbital frequency of $\Omega_r/(2 \pi) = 0.14331 \mathrm{mHz}$. The small yellow circle at the center represents the minimum periastron of the innermost stable orbit, which is $r_{\mathrm{ISO}}=p_{\mathrm{ISO}}/(1+e)$.}
		\label{SD_Geodesic}
	\end{figure}
	
	In Figure \ref{SD_Geodesic}, we can see that the three methods for obtaining the orbit essentially overlap. The equations \eqref{psifunel}-\eqref{gammafunel} or equations \eqref{psifunes} -\eqref{gammafunes} provide an approximate orbit, which accumulates errors over time. The approximation methods used in equations \eqref{psiApprox} and \eqref{phiApprox} introduce periodic errors between the actual orbit and the numerically solved geodesic orbit.
	
	By comparing the obtained orbit with the numerically solved orbit, we can determine that the approximate equations \eqref{psifunel}-\eqref{gammafunel} or equations \eqref{psifunes} -\eqref{gammafunes} we used are reliable. For further error analysis, please refer to Appendix \ref{OrbitError}.
	
	\subsection{Adiabatic evolution of frequency and eccentricity}
	We adopt the following approximation for the Keplerian orbital frequency $\nu$.
	\begin{eqnarray}
	 p = \left(\frac{1-P}{1+P}\right)^{2 / 3} \frac{X}{(2 \pi M \nu)^{2 / 3}}.
	\end{eqnarray}
	Combining equations \eqref{ETop}, \eqref{LTop}, \eqref{dEdt}, and \eqref{dLdt}, we can obtain the leading-order equations for the evolution of orbital frequency $\nu$ and eccentricity $e$.
	\begin{eqnarray}
		\frac{d \nu}{d t}&=&\frac{\mu^{2}}{M^{3} m} \frac{(1-P)^{6}}{(1+P)^{6}} \frac{(2 \pi M \nu)^{11 / 3}}{\left(1-e^{2}\right)^{7 / 2}} \frac{1}{10 \pi}\left(96+292 e^{2}+37 e^{4}\right)+\mathcal{O}((2 \pi M \nu)^{13/3}), \\
		\frac{d e}{d t}&=&-\frac{\mu^{2}}{15 m M^{2}} \frac{(1-P)^{5}}{(1+P)^{6}} \frac{(2 \pi M \nu)^{8 / 3}}{\left(1-e^{2}\right)^{5 / 2}} \frac{1}{e}\left(304 e^{2}+121 e^{4}-\left(192+280 e^{2}-47 e^{4}\right)P\right)+\mathcal{O}((2 \pi M \nu)^{10/3}).
	\end{eqnarray}
	It should be noted that, as $e \rightarrow 0$, $\Omega_\phi= 2 \pi \nu_\phi \rightarrow 2 \pi \nu$, which means $\nu \approx \nu_\phi$. Similarly, as $e \rightarrow 1$, $\Omega_r= 2 \pi \nu_r \rightarrow 2 \pi \nu$, which means $\nu \approx \nu_r$. The detailed analysis can be found in Appendix \ref{FProblem}.
	
	If $P$ is a small quantity less than or equals to $(2 \pi M v)^{2/3}$, then the leading order correction for $P$ is equal to or less than $3.5PN$, the evolution of $\nu$ and $e$ become: 
	\begin{eqnarray}
		\frac{d \nu}{d t}&=&\frac{\mu^{2}}{M^{3} m} \frac{(1-P)^{6}}{(1+P)^{6}} \frac{(2 \pi M \nu)^{11 / 3}}{\left(1-e^{2}\right)^{7 / 2}} \frac{1}{10 \pi}\left(96+292 e^{2}+37 e^{4}\right) \notag \label{nufun}\\
		&&+\frac{96 m}{10 \pi M^{3}} \frac{(2 \pi M \nu)^{13 / 3}}{\left(1-e^{2}\right)^{9 / 2}}\left(\frac{1273}{336}-\frac{2561}{224} e^{2}-\frac{3885}{128} e^{4}-\frac{13147}{5376} e^{6}\right)  +\mathcal{O}((2 \pi M \nu)^{5}) ,\\
		\frac{d e}{d t}&=&-\frac{\mu^{2}}{15 m M^{2}} \frac{(1-P)^{5}}{(1+P)^{6}} \frac{(2 \pi M \nu)^{8 / 3}}{\left(1-e^{2}\right)^{5 / 2}} e\left(304 +121 e^{2}-\left(192+280 e^{2}-47 e^{4}\right) \frac{P}{e^2}\right) \notag\\
		&& -\frac{m}{15 M^{2}} \frac{(2 \pi M \nu)^{10 / 3}}{\left(1-e^{2}\right)^{7 / 2}}e\left(\frac{1}{56}\left(70648-231960 e^{2}-106523 e^{4}\right)\right)  +\mathcal{O}((2 \pi M \nu)^{4}) . \label{efun}
	\end{eqnarray}
	Through Eq. \eqref{efun}, we can see that the leading order effect of the $P$ is positive. This means that the correction of LQG tends to increase the orbital eccentricity. As the eccentricity decreases, this effect becomes more significant. Moreover, when the eccentricity approaches 
	  \begin{eqnarray}
	  	e_B= 2 \sqrt{\frac{-38+35 P+\sqrt{1444-1208 P+1789 P^{2}}}{121+47 P}} \label{ePErr},
	  \end{eqnarray}
	 the leading-order effect of $P$ will reach the same order as that in the Schwarzschild case. At this point, for the evolution Eq. \eqref{efun}, the precision will degrade to the $2.5PN$ order.
	
	\subsection{Waveforms}
	In this paper, based on equations \eqref{psiApprox} and \eqref{phiApprox}, we use the quadrupole waveform \cite{Kludge_Babak_2007,Gravitational_Maggiore_2007} as follows:
	\begin{eqnarray}
		h^{jk} = \frac{2}{r} \ddot{\mathcal{Q}}^{jk}.
	\end{eqnarray}
	
	In the previous section, we assumed that the orbit is located in the equatorial plane, which means that the orbital angular momentum direction is $(\theta_L=0, \phi_L=0)$. Combining equations \eqref{QM}, \eqref{dtdtau}, \eqref{dphidtau}, and performing a Taylor series expansion, the leading-order gravitational waveform can be expressed as:
	
	\begin{eqnarray}
		h^{11}&=&\mathcal{A}_h\Big(2 e^{2}+2 e \cos(\psi(t))-5 e \cos(\psi(t)-2 \phi(t))\Big.\notag\\
		&&\Big.-2 e^{2} \cos(2 \psi(t)-2 \phi(t))-4 \cos(2 \phi(t))-e \cos(\psi(t)+2 \phi(t))\Big),\\
		h^{22}&=&\mathcal{A}_h\Big(2 e^{2}+2 e \cos(\psi(t))+5 e \cos(\psi(t)-2 \phi(t))\Big.\notag\\
		&&\Big.+2 e^{2} \cos(2 \psi(t)-2 \phi(t))+4 \cos(2 \phi(t))+e \cos(\psi(t)+2 \phi(t))\Big),\\
		h^{12}&=&\mathcal{A}_h\Big(5 e \sin(\psi(t)-2 \phi(t))+2 e^{2} \sin(2 \psi(t)-2 \phi(t))\notag\\
		&&-4 \sin(2 \phi(t))-e \sin(\psi(t)+2 \phi(t))\Big),\\
		h^{13}&=&h^{23}=h^{33}=0,
	\end{eqnarray}
	where, $\mathcal{A}_h={m(1-P)^{2}}/{(2p(1+P)^{2})}$.
	In the actual calculation, we computed up to the second order with $p$. 
	The traceless gauge of the waveform is given by\cite{Kludge_Babak_2007,Gravitational_Maggiore_2007}:
	\begin{eqnarray}
		h_{\mathrm{TT}}^{j k}=\frac{1}{2}\left(\begin{array}{ccc}
			0 & 0 & 0 \\ 
			0 & h^{\theta_S \theta_S}-h^{\phi_S \phi_S} & 2 h^{\theta_S \phi_S} \\ 
			0 & 2 h^{\theta_S \phi_S} & h^{\phi_S \phi_S}-h^{\theta_S \theta_S}
		\end{array}\right),
	\end{eqnarray}
	where
	\begin{eqnarray}
		h^{\theta_S \theta_S}&=&\cos ^{2} \theta_S\left(h^{11} \cos ^{2} \phi_S+h^{12} \sin 2 \phi_S+h^{2 2} \sin ^{2} \phi_S\right), \\ 
		h^{\theta_S \phi_S }&=&\cos \theta_S\left(-\frac{1}{2} h^{11} \sin 2 \phi_S+h^{12} \cos 2 \phi_S+\frac{1}{2} h^{2 2} \sin 2 \phi_S\right), \\
		h^{\phi_S \phi_S}&=&h^{11} \sin ^{2} \phi_S-h^{12} \sin 2 \phi_S+h^{2 2} \cos ^{2} \phi_S .
	\end{eqnarray}
	The polarized gravitational waveforms correspond to:
	$h_+(t)=h^{\theta_S \theta_S}-h^{\phi_S \phi_S}$ and $h_\times(t) = 2 h^{\theta_S \phi_S}$.

	\section{Results of Waveforms}\label{GW}
	
	\subsection{Waveforms under eS method}
	
	In calculations, we primarily used the ordinary differential equations (ODEs) given by Eqs. \eqref{psifunes}-\eqref{gammafunes} up to $(11,6)$ orders together with \eqref{nufun} and \eqref{efun}.

	\begin{figure}[!htb]
		\centering
		\subfigure[{$e=0.05$}]{\label{SD_1mHz_M10p6_e005}
			\includegraphics[width=0.45\textwidth]{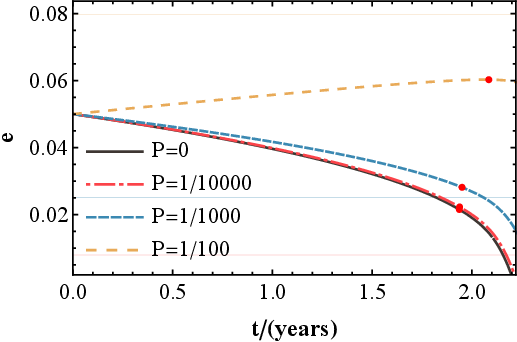}
			\includegraphics[width=0.45\textwidth]{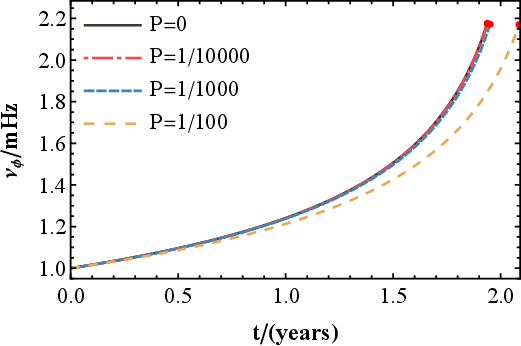}
		}
		\subfigure[{$e=0.1$}]{
			\includegraphics[width=0.45\textwidth]{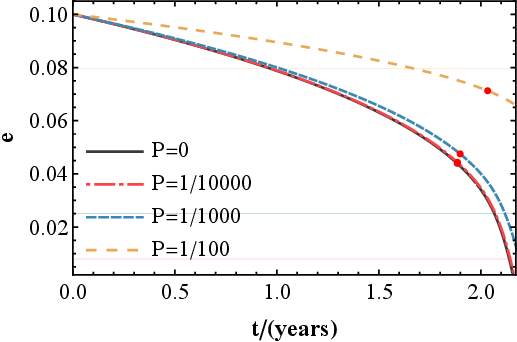}
			\includegraphics[width=0.45\textwidth]{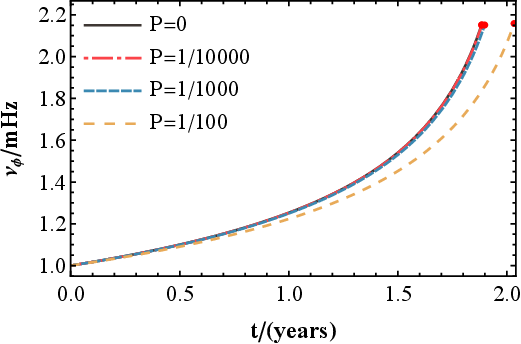}
		}
		\subfigure[{$e=0.2$}]{
			\label{SD_1mHz_M10p6_P_e005}
			\includegraphics[width=0.45\textwidth]{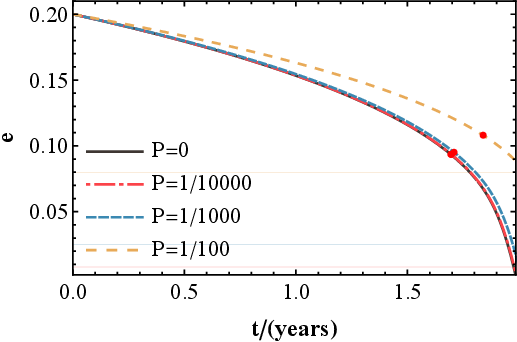}
			\includegraphics[width=0.45\textwidth]{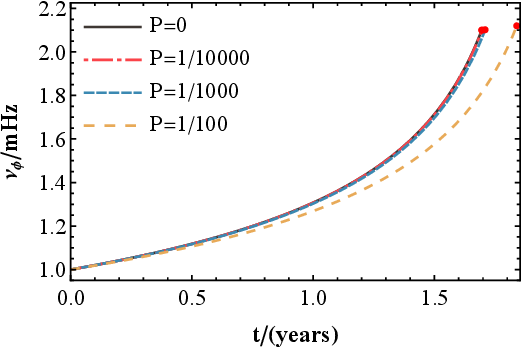}
		}
		\caption{The evolution of $e$ and $\nu_\phi$ for different initial eccentricities $(e_0=0.2, 0.1, 0.05)$ and an initial frequency of $\nu_\phi=1 \mathrm{mHz}$. The different horizontal lines in the background represent the critical eccentricity $e_B$. The red dot indicates the position of ISO.}
		\label{SD_1mHz_M10p6_e_P}
	\end{figure}
	
	As shown in Fig. \ref{SD_1mHz_M10p6_e_P}, by numerically solving equations \eqref{nufun} and \eqref{efun}, we can obtain the evolution of eccentricity $e$ and frequency $\nu_\phi$ with time. In the figure, the red dot represents the orbit reaching ISO, where the ISO is calculated in Appendix \ref{Append2}. It can be observed that with increasing $P$, both the decay rate of $e$ and the growth rate of $\nu_\phi$ decrease. Additionally, as shown in Fig. \ref{SD_1mHz_M10p6_e005}, when eccentricity $e$ is smaller than the limit $e_B$ in Eq. \eqref{ePErr}, the leading order term in Eq. \eqref{efun} becomes positive. When the leading order dominates, the eccentricity gradually increases during evolution. It is important to note that when the eccentricity evolves close to or below the dashed line, the accuracy of eccentricity evolution changes from $3.5PN$ to $2.5PN$.
	On the other hand, with increasing $P$, The growth rate of $\nu_\phi$ slows down. This implies that the quantum effect of the inspiraling binary plays a role similar to repulsion(reducing the impact of massive black holes).

	\begin{figure}[!htb]
		\centering
		\subfigure[{$e_0=0.05$}]{\label{SD_1mHz_M10p6_P_e01_psi}
		\includegraphics[width=0.45\textwidth]{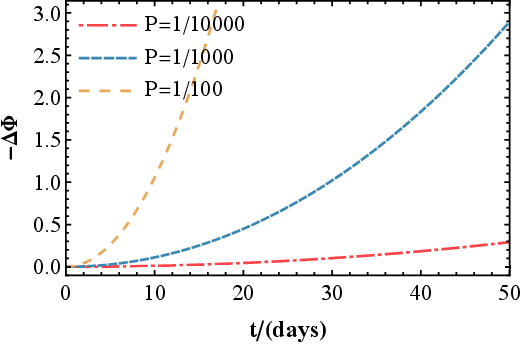}}
		\subfigure[{$P=1/1000$}]{\label{SD_1mHz_M10p6_P0001_e_phi}
		\includegraphics[width=0.45\textwidth]{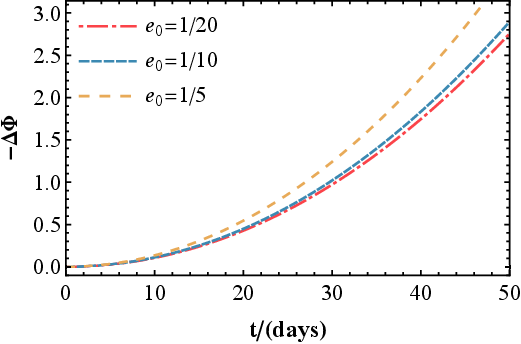}}
		\caption{The accumulated orbital phase difference $\Delta \Phi= \Phi(P)-\Phi(P=0)$ under different values of $P$ or $e_0$ with $M=10^6 M_\odot$ and $m=10 M_\odot$.}
		\label{SD_1mHz_M10p6_P_e005_psi}
	\end{figure}
	
	Fig. \ref{SD_1mHz_M10p6_P_e01_psi} shows the evolution of accumulated phase difference $\Delta \Phi = \Phi(P) - \Phi(P=0)$ with time for  various values of initial eccentricities $e_0$ and  $P$. It can be seen that with $e_0=0.1$, as $P$ increases, $\Delta \Phi$ increases faster. When $P=1/1000$, $\Delta \Phi$ grows more rapidly with increasing eccentricity, although the change is not significant.
	
\begin{figure}[!htb]
		\centering
		\subfigure[{$e_0=0.1$ , $P=1/10000$}]{
		\includegraphics[width=0.45\textwidth]{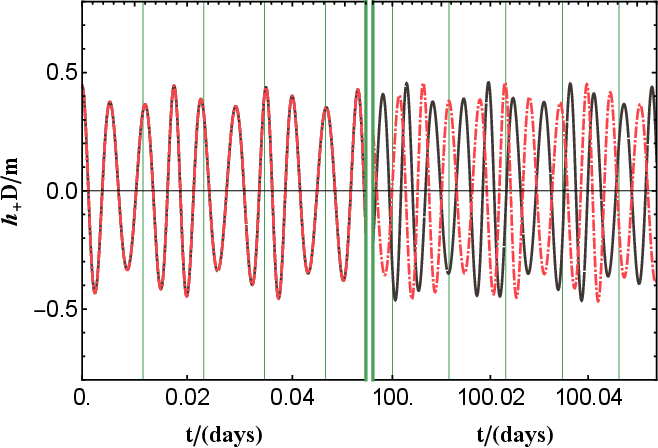}}
		\subfigure[{$e_0=0.2$ , $P=1/10000$}]{
		\includegraphics[width=0.45\textwidth]{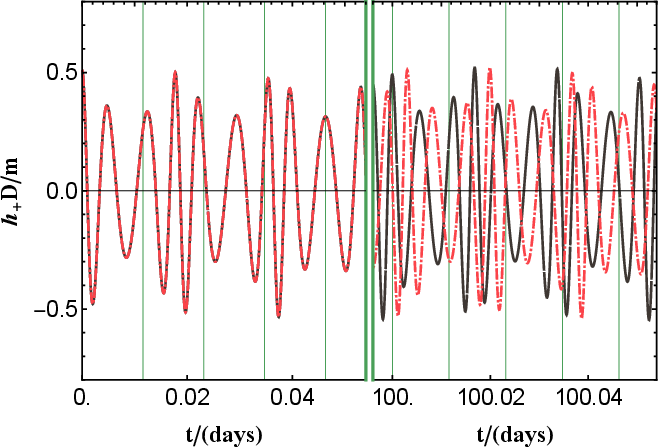}}
		\subfigure[{$e_0=0.1$ , $P=1/1000$}]{
		\includegraphics[width=0.45\textwidth]{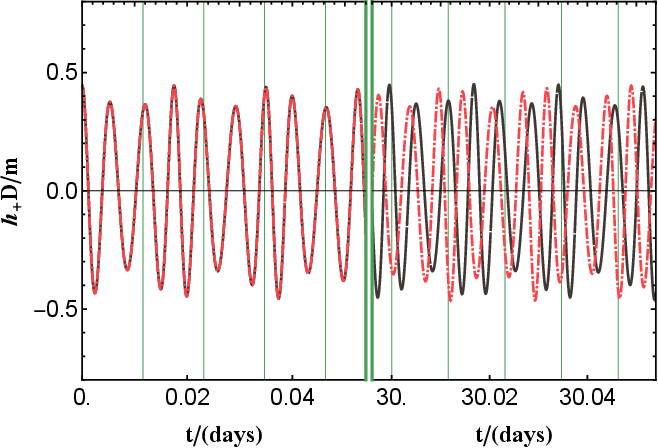}}
		\subfigure[{$e_0=0.2$ , $P=1/1000$}]{
		\includegraphics[width=0.45\textwidth]{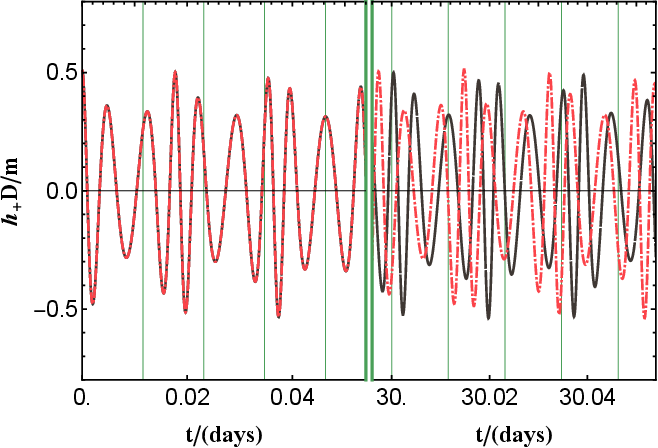}}
		\caption{The gravitational waveforms under  $e_0=\{1/10,1/5\}$ and  $P=\{1/10000,1/1000\}$. The solid black line represents the waveform with $P=0$. The left side of each plot shows the initial waveforms, while the right side shows the ones after a period of evolution.}
		\label{hPluseS}
	\end{figure}

	After performing the above calculations, we obtained the corresponding evolutions of $(\Psi, \nu_\phi, e_0, \gamma)$. Combining this with quadrupole waveform, we can obtain the gravitational waveforms for different values of $e_0$ and  $P$, as shown in Fig. \ref{hPluseS}. 
	
	In Fig. \ref{hPluseS}, it can be observed that when $e_0=0.1$ and $P=1/10000$, it takes around 100 days for the gravitational waveforms to show a significant difference. On the other hand, when $P=1/1000$, the difference becomes apparent in just 30 days. This is consistent with the conclusions in Fig. \ref{SD_1mHz_M10p6_P_e01_psi}.
	Similarly, when $P$ is fixed, the variation in eccentricity does not have a significant impact at which the gravitational waveforms exhibit noticeable phase differences.
	Moreover, each vertical line in the Fig. \ref{hPluseS} represents the time interval $T_\phi=1/\nu_\phi(t=0)$. We can find the GW frequency $\nu_{GW}(t=0)\approx 2 \nu_{\phi}(t=0)$, which is consistent with Ref. \cite{Improved_Chua_2015}.
	
	\subsection{Waveforms under eL method}
	The gravitational waveforms obtained through the large eccentricity method, described by equations \eqref{psifunel}-\eqref{gammafunel} up to $(10,5)$ orders with \eqref{nufun} and \eqref{efun}, can be used to generate waveforms for different values of  $e$ and $P$. 
	\begin{figure}[!htb]
		\centering
		\subfigure[{$e_0=0.6$ , $P=1/10000$}]{
			\includegraphics[width=0.45\textwidth]{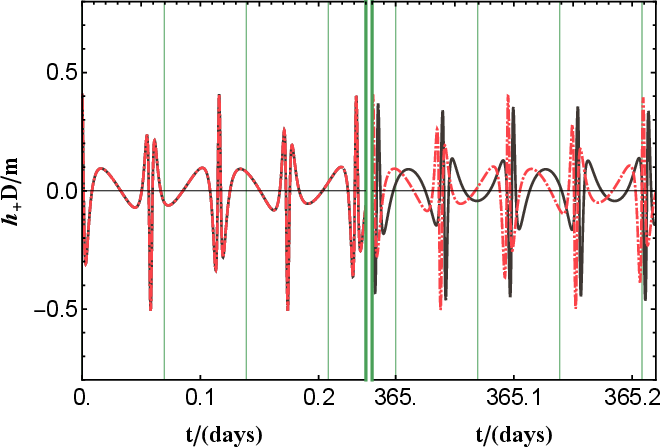}}
		\subfigure[{$e_0=0.7$ , $P=1/10000$}]{
			\includegraphics[width=0.45\textwidth]{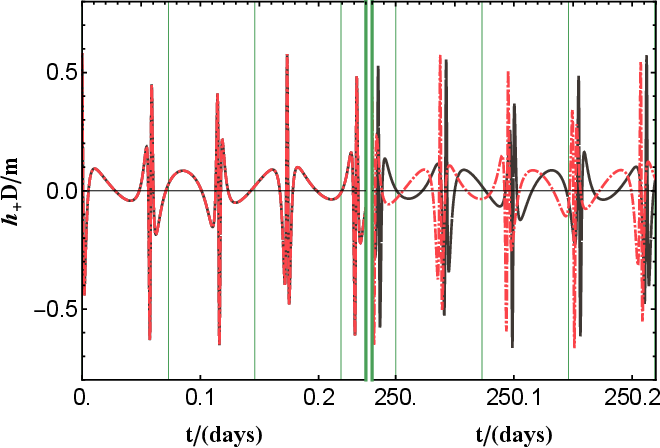}}
		\subfigure[{$e_0=0.6$ , $P=1/1000$}]{
			\includegraphics[width=0.45\textwidth]{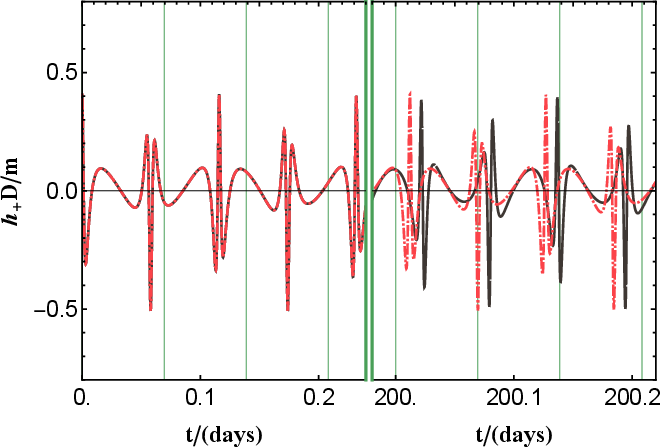}}
		\subfigure[{$e_0=0.7$ , $P=1/1000$}]{
			\includegraphics[width=0.45\textwidth]{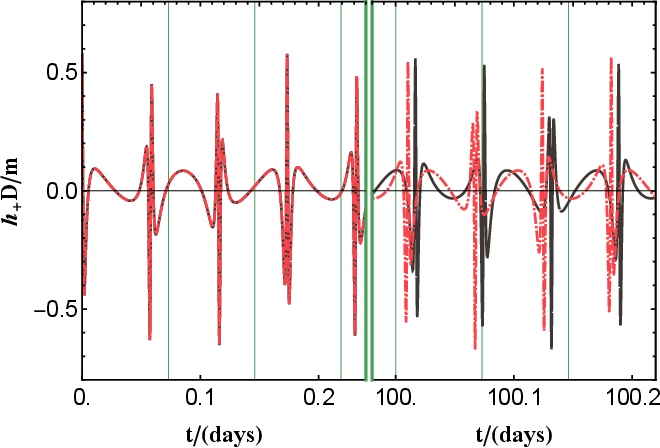}}
		\caption{The gravitational waveforms under different eccentricities $e_0=\{0.6,0.7\}$ and different loop quantum parameters $P=\{1/10000,1/1000\}$. The left side of each plot represents the initial waveforms, while the right side displays the ones after a period of evolution.}
		\label{hPluseL}
	\end{figure}
	Waveforms are illustrated in Fig. \ref{hPluseL}. It can be observed that when $e_0$ is fixed, a larger $P$ leads to a shorter time for significant differences to appear in the gravitational waveforms. Similarly, when $P$ is fixed, a larger $e_0$ results in a shorter time for noticeable differences in the waveforms. This trend is consistent with the evolution of $\Delta \Phi$ observed at smaller eccentricities. 
	
	\section{Parameter estimation}\label{PE}
	For a three arm spacebased GW detector, different sets of Michelson interferometers can be constructed by choosing different optical paths. The strain amplitude of the detector can be described by the unified expression:
	\begin{eqnarray}
		h(t)&=& \frac{1}{D}\frac{\sqrt{3}}{2} \left(F^+(t) h_+(t)+F^\times(t) h_\times(t) \right),
	\end{eqnarray}
	where the distance $D$ represents the separation between the observer and the GW source, while $F^+(t)$ and $F^\times(t)$ denote the antenna functions. 
	These functions depend on both the direction of the source $(\theta_S, \phi_S)$ and the orbital angular momentum direction $(\theta_L=0, \phi_L=0)$. The specific mathematical expressions can be found in references \cite{Spin-induced_Apostolatos_1994,LISA_Barack_2004}, where it is stated that these antenna functions have a period of one year ($T_{ap}=1 \mathrm{year}$), with initial parameter angles set to $\bar{\alpha}_0=\bar{\phi}_0=0$.
	Therefore, the GW signal is completely determined by the parameters 
	$\lambda=( M, m=10 M_\odot, P, \Phi_0 \approx 0, e_0, \nu_0(or \ t_0), \hat{\gamma}_0=0, \theta_S=\pi/2, \phi_S=\pi/2, \theta_L=0, \phi_L=0,  D= 1 Gpc)$
	where $e_0$ is the initial eccentricity of the orbit, and $\nu_0$ is the initial frequency of the orbit, corresponding to the chosen time range of the data. For example, in our data sampling, we select the data within $t_0=1 \mathrm{year}$ before the plunge $p_\text{plunge}=p+\delta p$, where a fixed shift $\delta p=0.1$ is applied. The corresponding $\nu_0$ will be obtained. 
	For the chosen source azimuthal angles $(\theta_S=\pi/2, \phi_S=\pi/2)$, this implies that $h_+=h^{11}$ and $h_\times =0$.
	In the subsequent calculations, we mainly consider the varying parameters $(M, P, e_0)$, while the remaining parameters are set to the aforementioned values.
	 
	To assess the effects of  the varying parameters,  the noise weighted inner product between two templates need to be introduce as \cite{Measuring_Flanagan_1998,Model_Lindblom_2008}:
	\begin{eqnarray}
		\langle h_{e} \mid h_{m}\rangle=2 \int_{0}^{\infty} \frac{h_{e}^{*}(f) h_{m}(f)+h_{e}(f) h_{m}^{*}(f)}{S_{n}(f)} d f,
	\end{eqnarray}
	where $h_{e}(f)$ and $h_{m}(f)$ are the frequency domain results of the $h_{e}(t)$ and $h_{m}(t)$ after Discrete Fourier transformation(DFT), respectively. $*$ denotes complex conjugation. $S_{n}(f)$ is the noise power spectral density(PSD) of the space-borne GW detector, such as LISA \cite{Laser_Amaro-Seoane_2017,Detecting_Maselli_2022}. Note that a Tukey window function with a cosine fraction of $0.01$ is used before DFT and the sampling frequencies of the GW signals are analyzed in Appendix \ref{Append3}. 
	
	$\rho$ represents the signal-to-noise ratio (SNR) of a signal, given by $\rho=\sqrt{\langle h \mid h\rangle}$.
	We can define faithfulness as:
	\begin{eqnarray}
		\mathcal{F}\left[P_e,P_m\right]= \frac{\left\langle h_{P=P_e} \mid h_{P=P_m}\right\rangle}{\sqrt{\left\langle h_{P=P_e} \mid h_{P=P_e}\right\rangle\left\langle h_{P=P_m} \mid h_{P=P_m}\right\rangle}}.
	\end{eqnarray}
	The corresponding mismatch is defined as:
	\begin{eqnarray}
		\mathcal{O}\left[P_e,P_m\right]=1-\mathcal{F}\left[P_e,P_m\right].
	\end{eqnarray}
	When the mismatch $\mathcal{O}$ is larger than $\sim \mathcal{D}/(2 \rho^2)$ ($\mathcal{D}$ represents the parameter space of the model, which is approximately $10$), it indicates significant differences between two GW signals and they cannot faithfully describe each other. For a signal with an SNR of 30, this requires a mismatch $\mathcal{O} \gtrsim 0.012$ ($\log_{10} \mathcal{O} \lesssim -1.92$) \cite{Measuring_Flanagan_1998,Model_Lindblom_2008}.
	
	\begin{figure}[!htb]
		\centering
		\subfigure[{$e_0=0.05$}]{\label{SD_Faithfulness_e005_M_P}
			\includegraphics[width=0.45\textwidth]{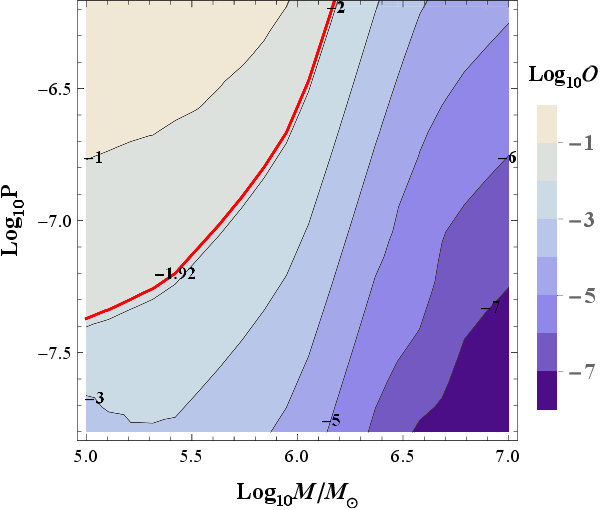}}
		\subfigure[{$e_0=0.1$}]{\label{SD_Faithfulness_e01_M_P}
			\includegraphics[width=0.45\textwidth]{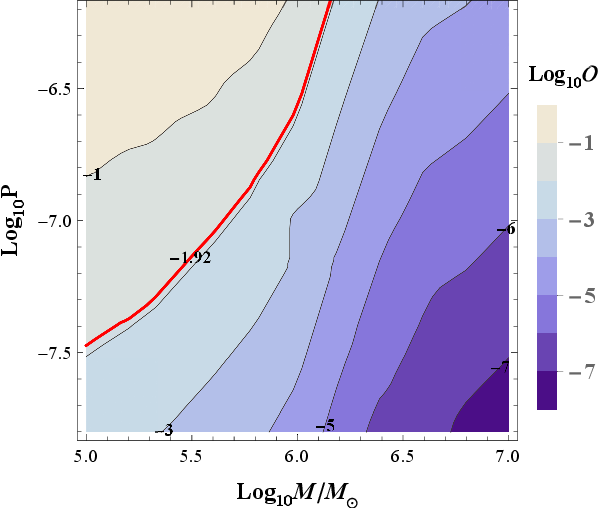}}
		\caption{ Mismatch of LISA  under different values of $M$ and $P$. The red line represents a boundary where mismatch $\mathcal{O} = 0.012$ ($\log_{10}\mathcal{O} = -1.92$).}
		\label{SD_Faithfulness_M_P}
	\end{figure}
	
	Fig. \ref{SD_Faithfulness_M_P} shows the constraint capability of the LISA detector on $P$ for eccentricities $e_0={0.05, 0.1}$ and different central black hole masses. From these figures, it can be seen that for one year of GW detection, the constraint capability of LISA on $P$ increases as the mass decreases within the range of $M=1 \times 10^5 M_\odot \sim 1 \times 10^7 M_\odot$. This is because as $M$ decreases from $1 \times 10^7 M_\odot$ to $1 \times 10^5 M_\odot$, the frequency $\nu_\phi$  around ISO  goes from about  $\sim 0.2 \mathrm{mHz}$ to $\sim 20 \mathrm{mHz}$, which is close to the LISA detector's most sensitive frequency $\sim 10 \mathrm{mHz}$. At the same time, the energy and angular momentum \textcolor{red}{flux(fluxes?)} increase as $M$ decreases.
	
	\begin{figure}[!htb]
		\centering
		\subfigure[{$M=10^5 M_\odot$}]{\label{SD_Faithfulness_M1t10p5}
			\includegraphics[width=0.45\textwidth]{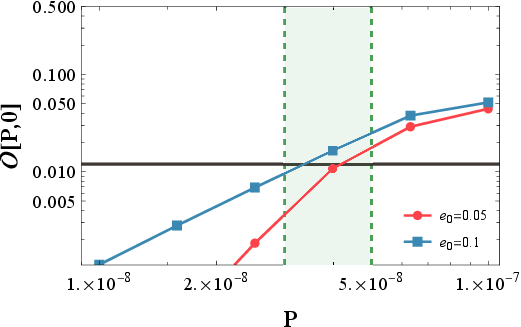}}
		\subfigure[{$M=10^6 M_\odot,$}]{\label{SD_Faithfulness_M1t10p6}
			\includegraphics[width=0.45\textwidth]{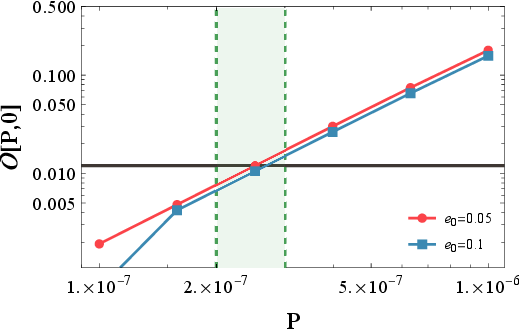}}
		\caption{Mismatch as a function of $P$ with different values of $e$ for LISA.The Black line represents a boundary where mismatch $\mathcal{O} = 0.012$ }
		\label{SD_Faithfulness}
	\end{figure}
	Fig. \ref{SD_Faithfulness} presents mismatch as a function of $P$ with different values of $e$ for LISA, in more detail. From Fig. \ref{SD_Faithfulness_M1t10p5}, it can be seen that for one year of GW detection with $M=1 \times 10^5 M_\odot$, LISA achieves a constraint capability of $3\times10^{-8} \sim 5\times10^{-8}$. When $M=1 \times 10^6 M_\odot$, as shown in Figure \ref{SD_Faithfulness_M1t10p6}, LISA achieves a constraint capability of $2\times10^{-7} \sim 3\times10^{-7}$.
	
	\section{Conclusion}\label{conclusion}
	
	In this paper, we consider the modified AK waveform in the EMRI system under the background of self-dual spacetime in loop quantum gravity. First, we calculate the motion equations of the test body in a self-dual spacetime and obtain the two orbital frequencies on the equatorial plane in the weak field limit. Then, combining the energy and angular momentum flux carried by the gravitational waves under the quadrupole moment approximation, we obtain the adiabatic orbit of the test body. Using different fundamental frequencies $\nu_r$ or $\nu_\phi$, we get two methods applied to different values of $e$.
	
	Through numerical calculations, we found that slight perturbations of $P$ will be amplified after one year of orbit evolution, and the final waveform results are significantly different from those without the  perturbation of $P$. 
	Among them, the smaller the mass $M$ of the central black hole, the stronger the constraint ability of LISA on the loop quantum parameters, in the range of $M \in [1 \times 10^5 M_\odot, 1 \times 10^7 M_\odot]$. Specially, the ability to constrain  $P$ will be approached the upper limit of about $3 \times 10^{-8} \sim 5 \times 10^{-8}$ with $M=1 \times 10^5 M_\odot$ for small changes in $e$ near zero. Overall, compared with the best constraint from the Cassini experiment in the solar system, $P\leq 5.5\times 10^{-6}$ \cite{Observational_Zhu_2020}, the constraint capability will be improved by 1 to 2 orders of magnitude.

    We only consider the self-dual spacetime in this paper, however, the methods described in the current paper can also be generalised to other loop quantum black hole models \cite{Quantum_Ashtekar_2005,Loop_Modesto_2006,Quantum_Ashtekar_2018,Mass_Bodendorfe_2021,Loop_Zhang_2023} and obtain the constraint on the relevant quantum parameter. We would like to leave these interesting topics for future study.

	\begin{acknowledgments}
		This work is supported by National Natural Science Foundation of China with No. 12275087.
	\end{acknowledgments}
	
	\appendix
	\section{Error analysis}\label{Append1}
	\subsection{Error of Frequency}\label{FProblem}
	Note that when $P=0$, equations \eqref{Omegar} and \eqref{Omegaphi} reduce to the Schwarzschild case. We obtain:
	\begin{eqnarray}
		&&\Omega_{\phi}=\frac{X^{3 / 2}}{M p^{3 / 2}}+\left(\frac{3 X^{3 / 2}}{M}-\frac{3 X^{5 / 2}}{M}\right) \frac{1}{p^{5 / 2}}+\left(\frac{57 X^{3 / 2}}{4 M}-\frac{63 X^{5 / 2}}{4 M}-\frac{15 X^{3}}{2 M}+\frac{9 X^{7 / 2}}{M}\right) \frac{1}{p^{7 / 2}} \notag\\ 
		&&+\left(\frac{315 X^{3 / 2}}{4 M}-\frac{96 X^{5 / 2}}{M}-\frac{60 X^{3}}{M}+\frac{237 X^{7 / 2}}{4 M}+\frac{45 X^{4}}{M}-\frac{27 X^{9 / 2}}{M}\right) \frac{1}{p^{9 / 2}} + \mathcal{O}(\frac{1}{p^{11/ 2}}), \label{OmegaphiS}\\
		&&\Omega_{r}=\frac{X^{3/2}}{M p^{3 / 2}}-\frac{3 X^{5 / 2}}{M p^{5 / 2}}  +\left(-\frac{6 X^{5/2}}{M}-\frac{15 X^{3}}{2 M}+\frac{9 X^{7 / 2}}{M}\right) \frac{1}{p^{7 / 2}}\notag \\ 
		&&+\left(-\frac{24 X^{5 / 2}}{M}-\frac{75 X^{3}}{2 M}+\frac{30 X^{7 / 2}}{M}+\frac{45 X^{4}}{M}-\frac{27 X^{9 / 2}}{M}\right) \frac{1}{p^{9 / 2}}  + \mathcal{O}(\frac{1}{p^{11/ 2}}).\label{OmegarS}
	\end{eqnarray}
	
	Interestingly, even if we calculate $\Omega_{\phi}$ and $\Omega_{r}$ up to $\mathcal{O}(p^{-23/2})$, we can still obtain the following approximation:
	\begin{eqnarray}
		\Omega_\phi \rightarrow \frac{X^{3 / 2}}{M p^{3 / 2}} , \quad When  \quad e \rightarrow 0 (X\rightarrow 1) ;\\
		\Omega_r \rightarrow \frac{X^{3 / 2}}{M p^{3 / 2}} , \quad When  \quad e \rightarrow 1 (X\rightarrow 0).
	\end{eqnarray}

	\begin{figure}[!htb]
		\centering
		\subfigure[{$e=0.1$ , $M= 10^6 M_\odot$}]{
		\includegraphics[width=0.45\textwidth]{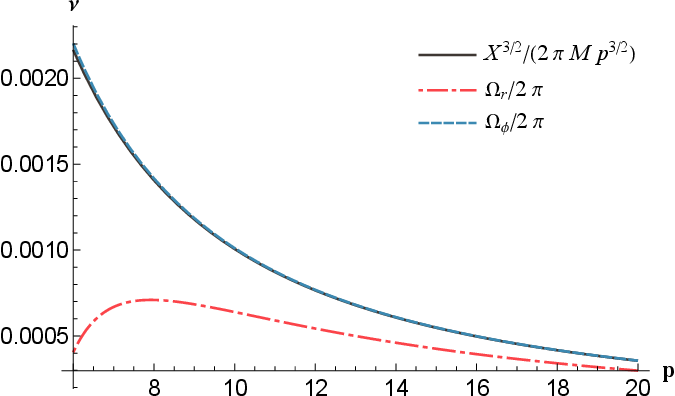}}
		\subfigure[{$e=0.8$ , $M= 10^6 M_\odot$}]{
		\includegraphics[width=0.45\textwidth]{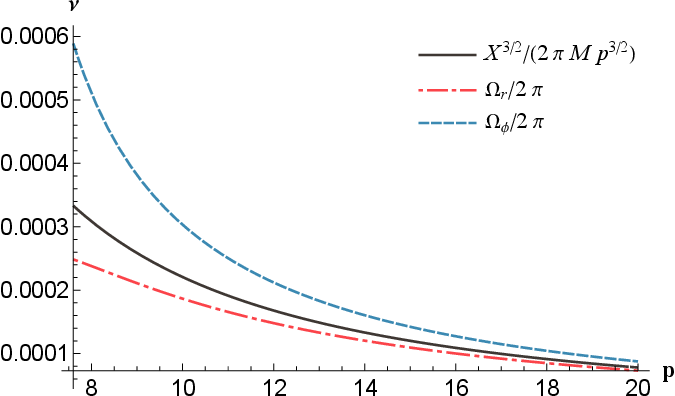}}
		\caption{ Relationship between the Kepler frequency $\nu={X^{3 / 2}}/{M p^{3 / 2}}$, the anomaly frequency $\Omega_r/(2\pi)$ and the ``longitude'' frequency $\Omega_\phi/(2\pi)$ with respect to $p$. Among them, $\Omega_r$ and $\Omega_\phi$ are both accurate to $\mathcal{O}(p^{-21/2})$ order.}
		\label{SelfDual_problem_nu}
	\end{figure}
	From the Fig. \ref{SelfDual_problem_nu}, we can see that when $e\rightarrow 0$, assuming $\Omega_\phi= 2 \pi \nu_\phi \approx \nu={X^{3 / 2}}/{M p^{3 / 2}}$ can achieve better accuracy.

	 When the parameter $P$ is small, it is only a small correction to the Eqs. \eqref{OmegaphiS}-\eqref{OmegarS} , as shown below:
	\begin{eqnarray}
	 	\Omega_\phi \rightarrow \frac{(1-P) X^{3/2}}{(1 + P) M p^{3/2}} + O(P) , \quad When  \quad e \rightarrow 0 (X\rightarrow 1) ;\\
	 	\Omega_r \rightarrow \frac{(1-P) X^{3/2}}{(1 + P) M p^{3/2}} + O(P) , \quad When  \quad e \rightarrow 1 (X\rightarrow 0).
	\end{eqnarray}
	 
	Therefore, we can assume the leading order to be 
	\begin{eqnarray}
		\Omega_\phi= 2 \pi \nu_\phi \rightarrow 2 \pi \nu, \quad When  \quad e \rightarrow 0 (X\rightarrow 1);\\
		\Omega_r= 2 \pi \nu_r \rightarrow 2 \pi \nu, \quad When  \quad e \rightarrow 1 (X\rightarrow 0).
	\end{eqnarray}
	
	In addition, when $e \rightarrow 0$, $\nu_r(p)$ is not a monotonic function outside the ISO, causing problems with the series expansion of its inverse function $p(\nu_r)$, which is a multivalued one. When $e \rightarrow 1$, the inverse series expansion $p(\nu_\phi)$ in Eq. \eqref{pTonu_phi} of $\nu_\phi(p)$ converges very slowly near the ISO.
	
	In summary, we have $\nu_\phi \approx \nu$ when $e\rightarrow 0$ and  $\nu_r \approx \nu$ while $e \rightarrow  1$.

	\subsection{Error of Geodesic orbit} \label{OrbitError}
	In the AK equation, when the energy flux and angular momentum flux tend to zero, the accumulated error of geodesic orbit is mostly concentrated in the accumulated error of $\hat{\gamma}$. Therefore, the error of the orbital approximation can be determine by Eqs. \eqref{psifunel}-\eqref{gammafunel} or Eqs. \eqref{psifunes}-\eqref{gammafunes} together with Eqs. \eqref{EquatorialEqT}-\eqref{EquatorialEqPhi}.
	
	Now, let us consider the case when
	$$ \{e=0.1(or \  0.8) ,\nu_\phi=0.2 mHz, M=10^6 M_\odot,P=1/1000\} $$
	
\begin{figure}[!htb]
		\centering
		\subfigure[{$e=0.1$}]{
			\includegraphics[width=0.45\textwidth]{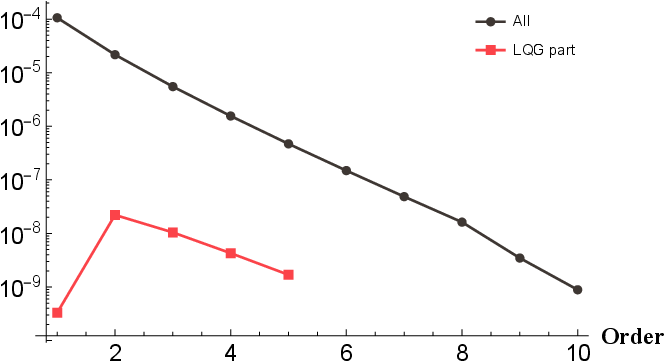}}
		\subfigure[{$e=0.8$}]{\label{SD_Error_gamma_eL08}
			\includegraphics[width=0.45\textwidth]{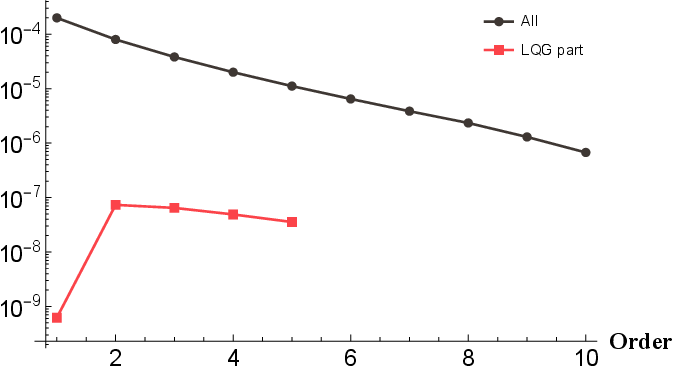}}
		\caption{Influence of different order under eL method.}
		\label{SD_Error_gamma_eL}
	\end{figure}

	Thus, $\dot{\hat{\gamma}}$ can be expanded in a series using $(2 \pi M \nu_\phi)^{2/3}$(or $(2 \pi M \nu_r)^{2/3}$), and the effects of different orders in the expansion are shown in Fig. \ref{SD_Error_gamma_eS} and \ref{SD_Error_gamma_eL}:
	\begin{figure}[!htb]
		\centering
		\subfigure[{$e=0.1$}]{
		\includegraphics[width=0.45\textwidth]{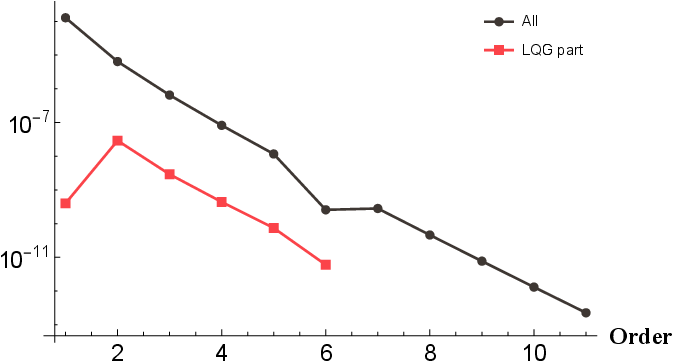}}
		\subfigure[{$e=0.8$}]{
		\includegraphics[width=0.45\textwidth]{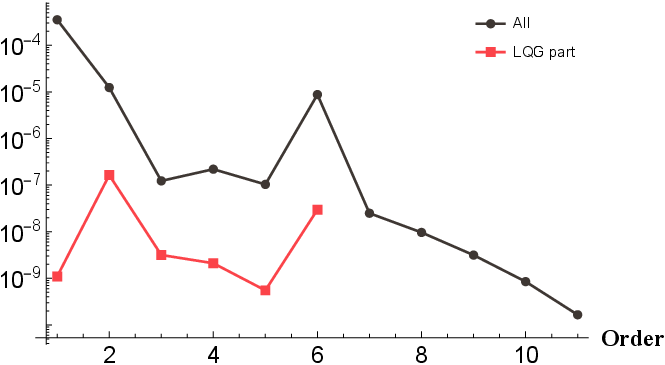}}
		\caption{Influence of different order under eS method.}
		\label{SD_Error_gamma_eS}
	\end{figure}
	
	The black dotted line represents the magnitude of each order in the series expansion of $\dot{\hat{\gamma}}$. The red square dotted line represents the magnitude of each order caused by the LQG effect. It can be seen that both methods achieve high accuracy. The convergence rate of the eS method  is faster overall, while the eL one is more stable.
	
	\begin{figure}[!htb]
		\centering
		\subfigure[{$e=0.1$ and $M= 10^6 M_\odot$}]{\label{SD_Error_gamma_e01}
			\includegraphics[width=0.45\textwidth]{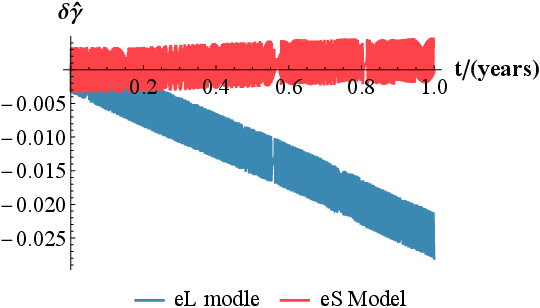}
		}
		\subfigure[{$e=0.8$ and $M= 10^6 M_\odot$}]{\label{SD_Error_gamma_e08}
			\includegraphics[width=0.45\textwidth]{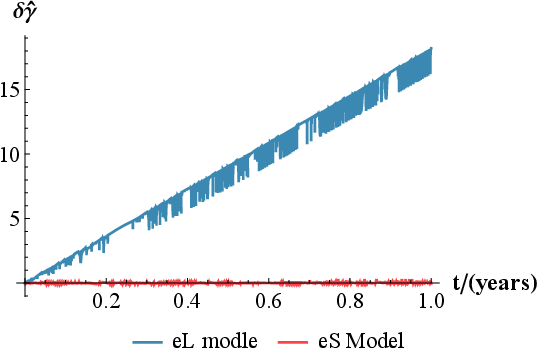}
		}
		\caption{ Error of $\hat{\gamma}$ for 1 year.}
		\label{SD_Error_gamma}
	\end{figure}
	In Fig. \ref{SD_Error_gamma}, the difference between the precessional angle , $\gamma(t)=\phi(t)-\psi(t)$, obtained by numerically solving Eqs. \eqref{EquatorialEqT}-\eqref{EquatorialEqPhi} and the approximate one in the AK method, $\gamma_{Ap}=\phi_{Ap}-\psi_{Ap}$ (see equations \ref{psiApprox} and \ref{phiApprox}), is shown as $\delta \hat{\gamma}$. 
	
	It can be seen from Fig. \ref{SD_Error_gamma_e01} that when $e \rightarrow 0$ such as $e=0.1$, after one year of evolution, the value of $\delta \hat{\gamma}$ is still much smaller than $1 \  \mathrm{rad}$, and the eS method is better than the eL one. 
	when $e \rightarrow 1$ such as $e=0.8$, the $\delta \hat{\gamma}$ from  eL method increases rapidly, which is due to the slow convergence as shown in Fig. \ref{SD_Error_gamma_eL08}. For example, the accuracy at the 10th order is approximately $5\times 10^{-7}$, which means that the error $\delta \hat{\gamma}$ after one year is approximately $5 \times 10^{-7} (\mathrm{year}/\mathrm{s}) \ \mathrm{rad} = 15 \mathrm{rad}$, consistent with Fig. \ref{SD_Error_gamma_e08}. This further confirms the previous conclusion. Note that the error is not a line but is caused by the periodic error of $\gamma_{Ap}$. In conclusion, when the initial eccentricity $e$ is relatively small, we apply the eS method, using $\nu_\phi$ as the fundamental frequency,  for the calculation and analysis of gravitational waveforms.
	
	\section{Innermost stable orbit}\label{Append2}
	Note that $a_0$ appears in the higher order terms and is a very small quantity that can be set to zero. To find the ISO, we can make  $d((dr/d\tau)^2)/dr=0$ at the pericenter $r_{\text{peri}}$\cite{Rotating_Bardeen_1972, Location_Stein_2020}. Combining equations \eqref{drdtau}, \eqref{Etope}, and \eqref{Ltope}, we obtain the separatrix equation

	\begin{small}
	\begin{eqnarray}\label{SeparatrixEq}
		0 &=&512 e (1 + e)^3 \bigl(3 + e (-2 + 3 e)\bigr) P^7 
		- 32 e (1 + e)^2 P^5 (1 + P)^2 \mathcal{S}_1 p \notag\\
		&&+ 16 e (1 + e) P^3 (1 + P)^4 \mathcal{S}_2 p^2
			+ 8 e (1 + e) P^2 (1 + P)^6 \mathcal{S}_3 p^3 \notag\\
		&&+ 4 e (1 + e) (-1 + P)^2 P (1 + P)^8 \Bigl(15 + 8 e + e^2 - 2 \bigl(5 + e (2 + e)\bigr) P + (3 + e) (5 + e) P^2\Bigr) p^4 \notag\\
		&&+ 2 e (1 + e) (-1 + P)^2 (1 + P)^{10} \bigl(3 + e -  (5 + e) P + (3 + e) P^2\bigr) p^5 \notag\\
		&&-  e (1 + e) (-1 + P)^2 (1 + P)^{12} p^6,
	\end{eqnarray}
	\end{small}
	where 
	\begin{small}
	\begin{eqnarray}
		\mathcal{S}_1&=&37 + 8 e^3 (-1 + P)^2 + e^4 (-1 + P)^2 - 16 e P + P (-50 + 37 P) + 6 e^2 \bigl(3 + P (-2 + 3 P)\bigr) ,\\
		\mathcal{S}_2&=&15 + e^4 (-1 + P)^2 \bigl(1 + (-1 + P) P\bigr) \notag\\
		&& + 2 e^3 (-1 + P)^2 \bigl(3 + P (-2 + 3 P)\bigr) - 6 e^2 P \bigl(5 + P (-6 + 5 P)\bigr) \notag\\
		&&+ P \Bigl(-71 + P \bigl(88 + P (-71 + 15 P)\bigr)\Bigr) + 2 e \biggl(5 + P \Bigl(-36 + P \bigl(38 + P (-36 + 5 P)\bigr)\Bigr)\biggr),\\
		\mathcal{S}_3&=&27 + e^3 (-1 + P)^4 + e^2 (-1 + P)^2 \bigl(1 + (-14 + P) P\bigr) + P \Bigl(-64 + P \bigl(98 + P (-64 + 27 P)\bigr)\Bigr) \notag\\
		&&+ e \biggl(19 + P \Bigl(-44 + P \bigl(74 + P (-44 + 19 P)\bigr)\Bigr)\biggr),
	\end{eqnarray}
	\end{small}
	Assuming $P$ is small and accurate up to second order of $P$, the approximate solution to the equation is obtained as
	\begin{eqnarray} \label{SeparatrixApprox}
	p = 6 + 2e - (12 + 4e) P + \frac{6 \bigl(35 + 35 e + 9 e^2 + e^3\bigr)}{(3 + e)^2}P^2.
	\end{eqnarray}

	\begin{figure}[H]
		\centering
		\includegraphics[width=0.45\linewidth]{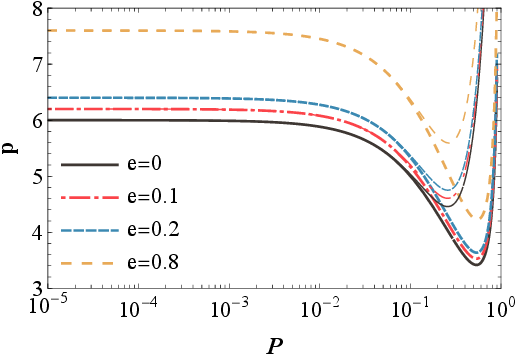}
		\caption{For different values of $e$, the $p$ of ISO respect to $P$, where the thick line is the numerical solution of Eq. \eqref{SeparatrixEq}, while  the thin lines are the approximate solutions of Eq. \eqref{SeparatrixApprox}.}
		\label{SD_M1t10p6_P_e_ISO}
	\end{figure}
	From the Fig. \ref{SD_M1t10p6_P_e_ISO}, when P is less than 0.01, the approximate solution can be obtained with high precision.
	For example, when the parameters are $\{P=1/100, e=0.2\}$, we take the approximate $p=6.27448$, and the corresponding numerical solution to Eq. \eqref{SeparatrixEq} is $p=6.27445$, with a relative error of approximately $5.8\times 10^{-6}$.

	\section{Discrete Fourier transform}\label{Append3}

	\begin{figure}[!htb]
		\centering
		\subfigure[{$h(f)$ when $e \rightarrow 0$, initial frequency $\nu_\phi=1\mathrm{mHz}$. The vertical line is $\nu=2 \nu_\phi+\mathbf{n}\nu_r, \mathbf{n}=0, \pm1,\pm2,...$ for $e=0.05$}]{
		\includegraphics[width=0.45\textwidth]{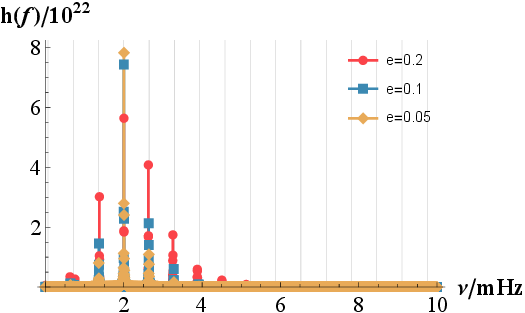}
		}
		\subfigure[{ $h(f)$ when $e \rightarrow 1$, initial frequency $\nu_\phi=0.2 \mathrm{mHz}$. The vertical line is $\nu=\mathbf{n}  \nu_\phi, \mathbf{n}=0, \pm1,\pm2,...$ for $e=0.6$}]{
		\includegraphics[width=0.45\textwidth]{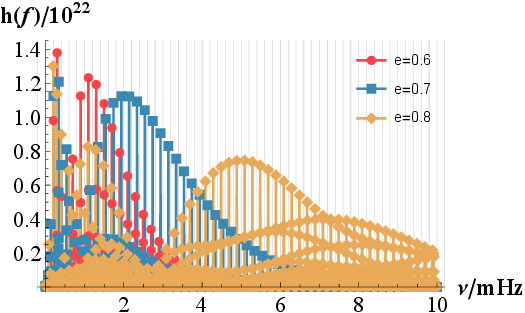}
		}
		\caption{ $h(f)$ from DFT of $h(t)$ for $e={0.05, 0.1, 0.2}$ with $P=0.001$.}
		\label{SD_eS_hIf}
	\end{figure}
	Fig. \ref{SD_eS_hIf} shows $h(f)$ from DFT of $h(t)$. It can be seen that as the eccentricity decreases, the contribution of the GW frequency $\nu=2 \nu_\phi+\mathbf{n}\nu_r, \mathbf{n}=\pm1,\pm2,...$ becomes more and more significant. In particular, the contribution of $\nu=2 \nu_\phi$ reaches its maximum value. This is consistent with the fact that as $e\rightarrow 0$, the GW frequency approaches $2 \nu_\phi$. 
	Moreover, for $e=0.1$, the contributions of frequencies $\nu=2 \nu_\phi + 2 \nu_r$ cannot be neglected.  When the signal is sampled, the sampling frequency can be selected as $f_{Ny} = 2 \nu_\phi+ 2 \nu_r < 4 \nu_\phi$.
	
	For example, in the case of $M= 10^6 M_\odot$, with an initial eccentricity of $e=0.2$, after 1 year of evolution to the innermost stable orbit, the eccentricity is still approximately $e\approx 0.1$ and corresponds to $\nu_\phi\approx 2 \mathrm{mHz}$ and $\nu_r\approx 0.5 \mathrm{mHz}$. At this point, the sampling frequency $f_{S} = 2.5 (2 \nu_\phi+ 2 \nu_r) \approx 12.5 \mathrm{mHz}$.

	\bibliographystyle{unsrt}

\end{document}